\documentclass[12pt,epsf,epsfig,psfig]{article}
\usepackage{graphicx}
\usepackage{epsfig}
\def\J{$J/\psi$}
\def\j{J/\psi}

\def\C{c{\bar c}}

\def\be{\begin{equation}}
\def\ee{\end{equation}}
\def\lsim{\raise0.3ex\hbox{$<$\kern-0.75em\raise-1.1ex\hbox{$\sim$}}}
\def\gsim{\raise0.3ex\hbox{$>$\kern-0.75em\raise-1.1ex\hbox{$\sim$}}}

\begin{document}



\centerline{\Large \bf Integers \& Prime Numbers: Deriving Zipf's Law}

\vskip1cm

\centerline{\large \bf Helmut Satz$^*$}

\medskip

\centerline{\bf Fakult\"at für Physik, Universit\"at Bielefeld, Germany}

\vskip 0.8cm

\centerline{\large \bf Abstract}

\bigskip

In a prime factor decomposition of integers in a given set, the occurrence 
frequencies of prime numbers are shown to satisfy a general form of Zipf's 
law. 

\vskip1cm


{\sl When the probability of measuring a particular value of some quantity 
varies inversely as a power of that value, the quantity is said to follow 
Zipf's law \cite{newman}}.

\medskip
 
 Around the middle of the past century,   
the Harvard linguist George Kingsley
Zipf had ranked words in texts in the English language
according to the frequency of their occurrence \cite{Zipf1,Zipf2}.
He denoted with 
$f(r=1)$ the number of times the most frequent word appeared, 
$f(r=2)$ for the next, and so on. The sequence started with the words 
``the, of, and, to, in, a, that, ...''. 
Zipf then observed that the frequency of the first rank, $f(r=1)$ for ``the'', 
was twice that of the second, $f(r=2)$ for ``of'', three times that of $r=3$, 
and so on. The resulting general relation 
\be
f(r)= {{\rm const.}\over r}
\ee
is denoted as Zipf's law; it states that the probability for rank $r$ of a 
given word varies inversely with $r$. The constant depends on the total
number of words in the text. In Fig.\ 1, we show a recent 
analysis \cite{jager}, based on
the SUSANNE corpus (128.000 words); apart from some fluctuations at both
ends, the agreement is good over four orders of magnitude.

\vskip0.1cm

\begin{figure}[h]
\hfill{\epsfig{file=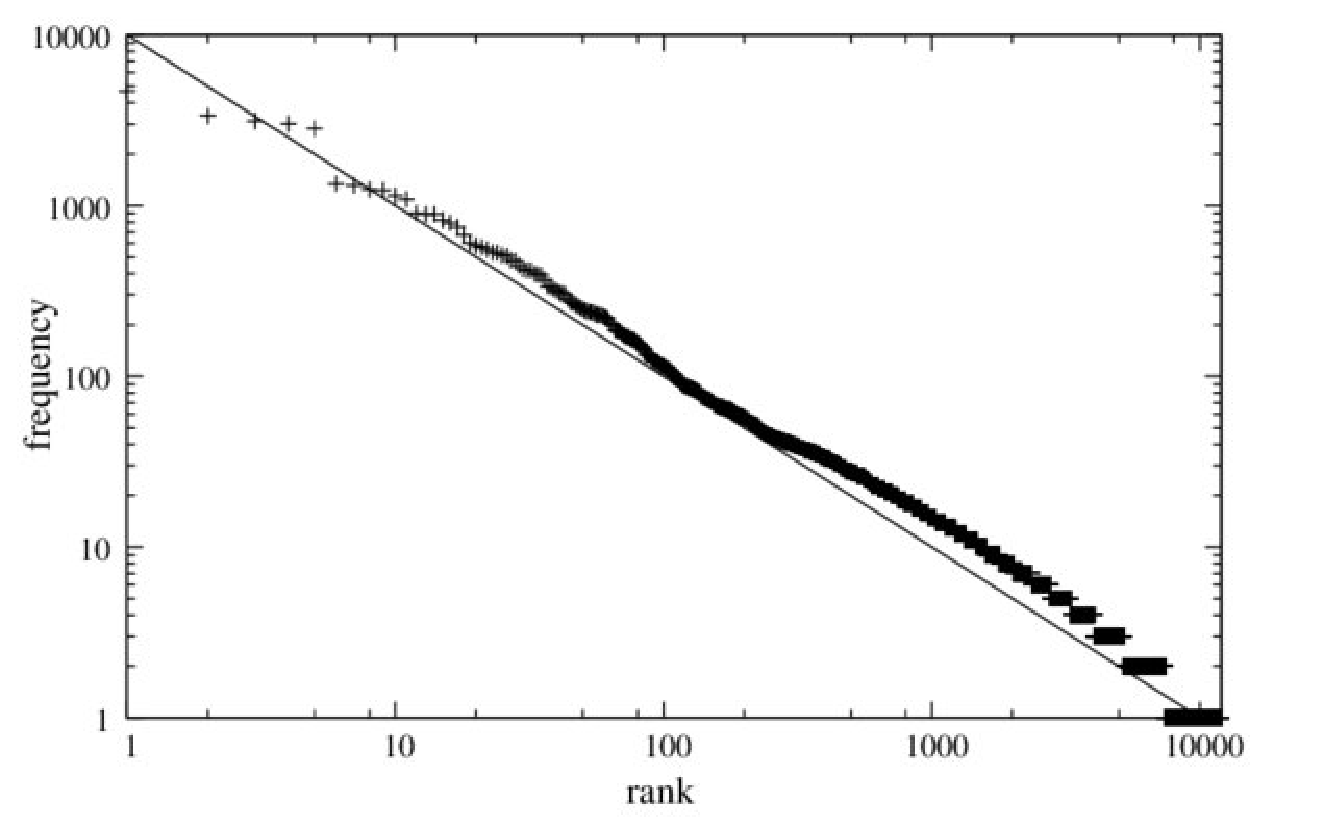,angle=-90,width=12cm}}
\end{figure}

\vskip-6cm

Fig. 1: 

\medskip

Word frequencies vs. 

rank in the English 

language \cite{jager}; solid 

line: Zipf's law.

\vfill

* satz@physik.uni-bielefeld.de

\newpage

The law was subsequently shown to hold
not only for most English texts, but for texts in practically all
other languages as well \cite{Shtrikman}, see Fig.\ 2, where it is shown in 
logarithmic form, $ \log f(r) = {\rm const.} - \log r$.
It is found to hold even for ancient languages not yet deciphered -
the meaning of the words there is not known, but their frequencies follow
Zipf's law. The law thus consists of two aspects: a partitioning of the text 
into subsets ordered according to the occurrence frequency of words, and the 
observation that the frequency distributions follow the ranking according 
to eq.\ (1). 

\vskip-0.5cm
\begin{figure}[h]
\centerline{\epsfig{file=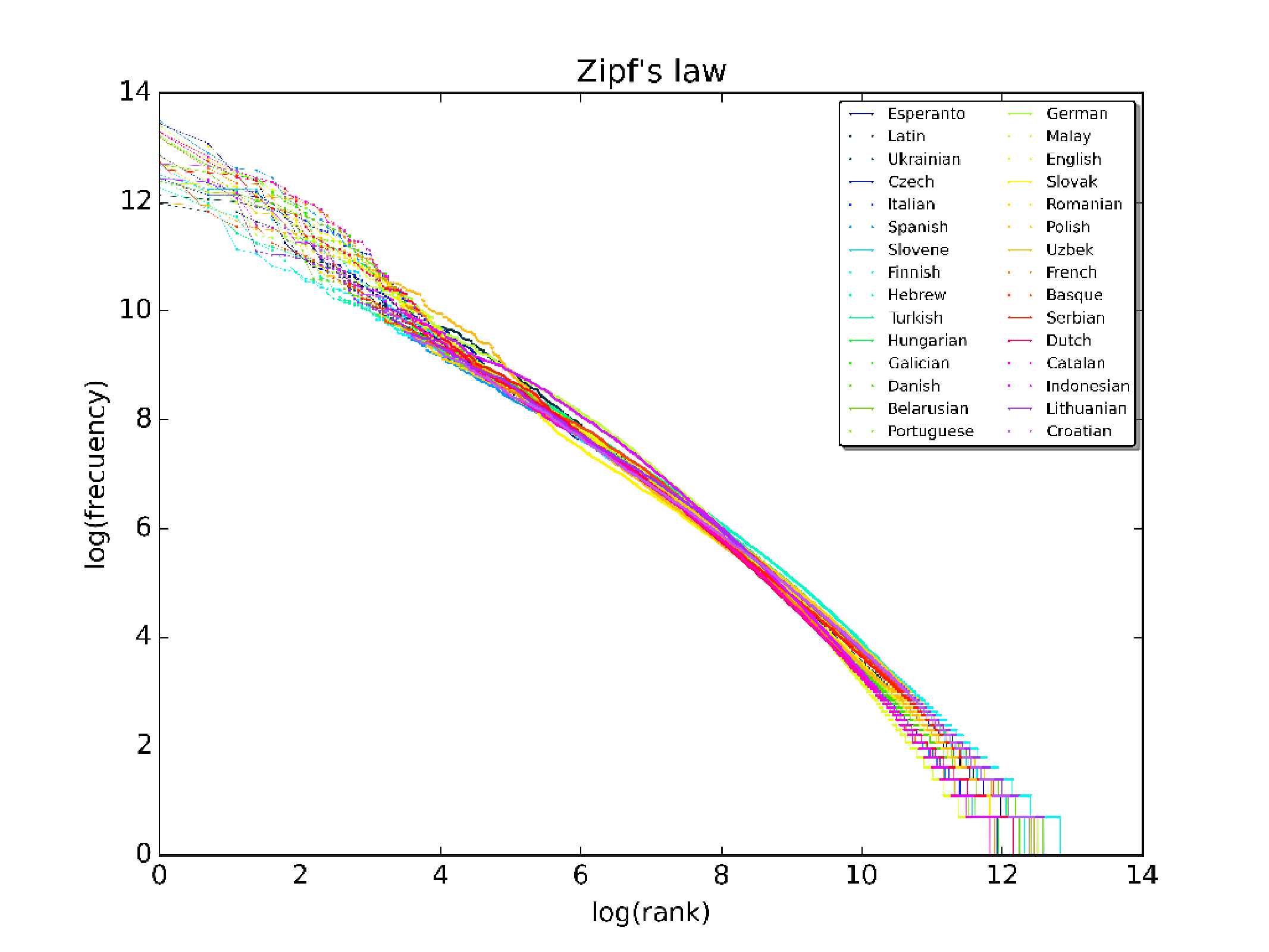,angle=-90,width=16.5cm}}
\end{figure}

\vskip-0.3cm

\centerline{Fig. 2: Word frequencies in different languages \cite{Shtrikman}}

\vskip0.5cm

We note that eq.\ (1) implies scale-invariance: increasing the rank by a
factor $x$ decreases the word frequency by a factor $1/x$, no matter what the
rank is. To allow for some 
deviations, Mandelbrot \cite{Mandelbrot} proposed a generalized form,
\be
f_m(r) \sim {1\over(r + b)^x},
.\ee
with constants $b$ and $x$. This allowed modifications near $r=1$, as well 
as a variation of the power $x=1$. For this more general form,  
scale-invariance holds only asymptotically in $r$. 

\medskip

In recent years it has become evident that a great number of very different
phenomena in very different areas of nature follow this universal 
law. Nevertheless, the origin of the law remains unknwown. Various attempts 
had suggested linguistic input elements - least efforts in communication, 
and more. It is also not a consequence of the length of the words; the
one-letter word ``a'' is six ranks below the three-letter word ``the''.
The validity of such explanation atttempts became even more doubtful 
with the realization that laws of this kind hold, as already indicated, for 
a great variety of phenomena in nature, quite detached from any human 
influence. We note only two instances. 

\medskip

The Gutenberg-Richter
law for the distribution of earthquakes follows eq.\ (1), with $m$ denoting the
magnitude of the quake (its logarithmic energy output) and $f(m)$ the frequency of occurrence for such events.
A devastating earthquake is no longer considered a unique ``singular'' event, 
but becomes one in a series
of ever increasing eruptions, according to a Zipf law, see Fig.\ 3. The
deviation for small quakes is a result of the difficulty to measure such events
prescisely. 

\begin{figure}[h]
\centerline{\epsfig{file=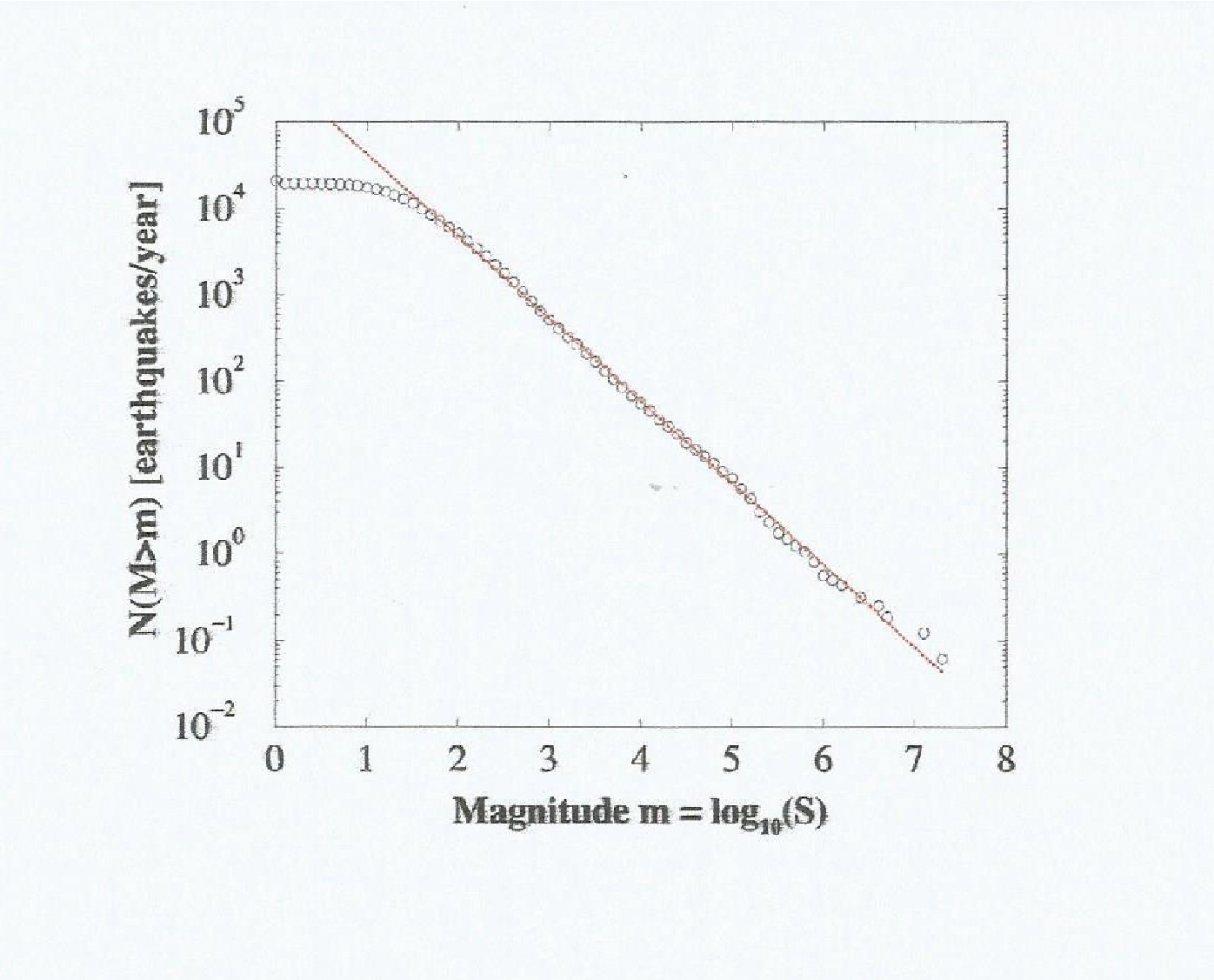,width=9cm,angle=-90}}
\end{figure}

\vskip-0.5cm

\centerline{Fig. 3: The number of earthquakes as function of their magnitude
\cite{Bak}}

\bigskip

Previous to Zipf, the German physicist Felix Auerbach \cite{Auerbach} had  
concluded already in 1913 that the distribution of city sizes, population 
vs.\ number of cities, follows a Zipfian pattern. This prediction became 
more firmly established \cite{Jiang}, once the size of cities was no longer 
defined by human zoning laws, but by the electric light they emitted 
at night - when New York became the shining greater New York, instead that 
of the legal districts; see Fig. 4.

\begin{figure}[h]
\centerline{\epsfig{file=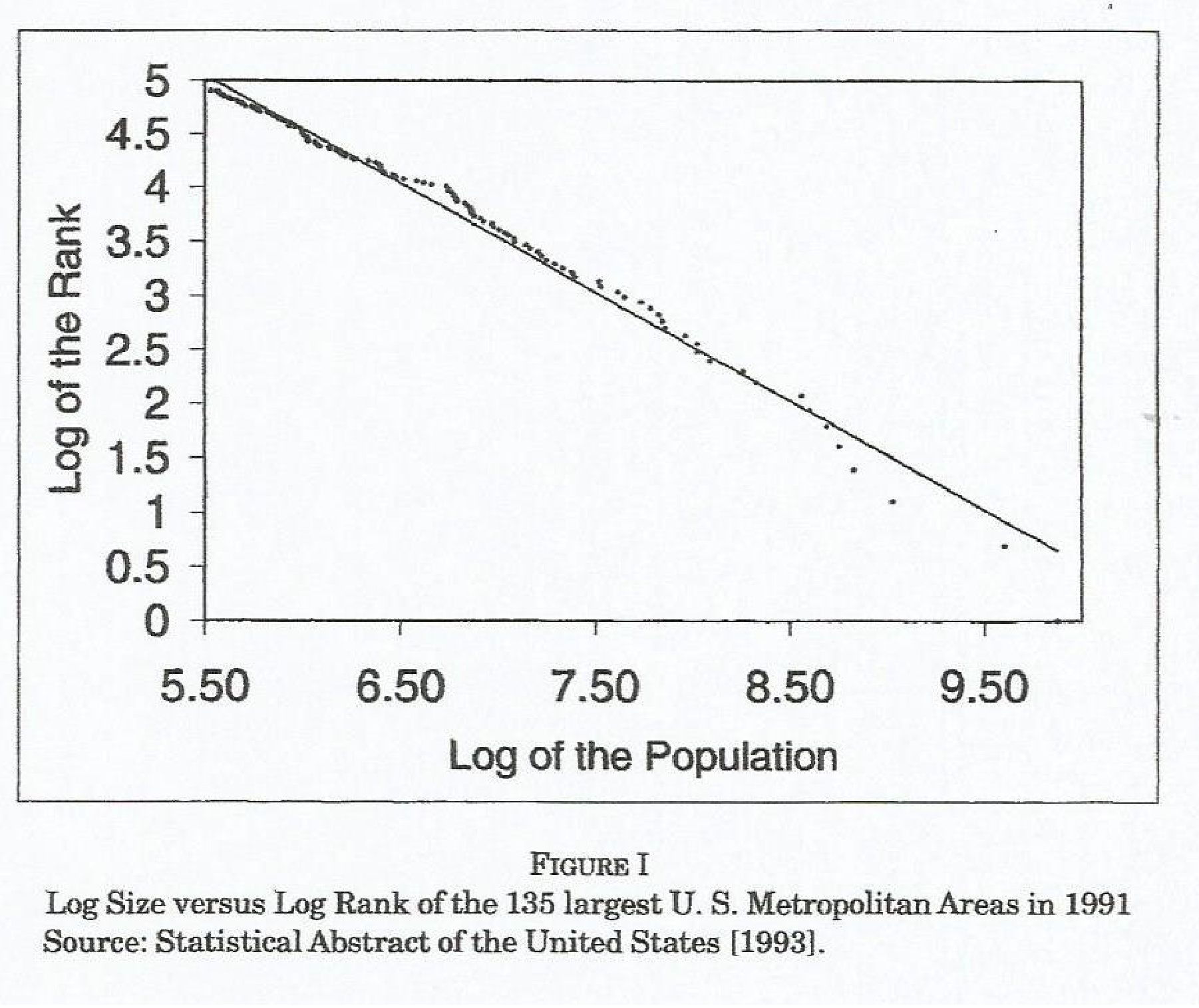,width=8cm,angle=-90}}
\medskip
\centerline{Fig. 4: City sizes  ranked vs. populations \cite{Jiang}}
\end{figure}

\medskip 

More recently, a great number of economic events were also found to 
follow such a pattern; for references, see \cite{newman}. It thus seems 
unlikely that the law arises because of some unknown linguistic human 
reasoning. 

\medskip

Therefore, to study a case quite detached from any human speculation, and
moreover one for which Zipf's law can be derived, rather than concluded by
observation, we turn to a linguistic analogue in mathematics. The integers, 
written in terms of their 
prime factors, provide possible ``texts'', containing the prime numbers as
``words''. In the following, we carry out such an analysis.
   
\medskip

We start for simplicity with the first one hundred 
integers 1,~2,~3,~4,...,~N=100. We write these in their prime number
decomposition, 2,~3, 4=2x2,~5,~6=2x3,~7,~8=2x2x2,~9=3x3, ~10=2x5, etc. 
As mentioned, the set of integers will be taken as our text and the prime 
numbers are the words in this text. Thus 90 = 2 x 3 x 3 x 5 consists of the 
words 2, 3, and 5. We now count how often a specific prime factor occurs in 
the given set of decomposed integers. To illustrate: for the first 10
integers, the two occurs eight times (2, 2x2, 2x3, 2x2x2,2x5), the three
four times (3,2x3,3x3), the five twice (5,2x5) and the seven once.

\medskip

Similarly, we find that in the first one hundred 
integers, 
the 2 occurs 97 times.  The 3 occurs 48 times, the 5 then 24 times, and so on; 
the results are listed in Table 1. In Fig. 5, we show the logarithmic form, 
$\log~f(N.p)$ vs. $\log~p$, for the
mentioned interval, and it is seen that Zipf's law is quite well satisfied.
The more general Mandelbrot form provides, as we shall see shortly, a really
ideal fit.

\medskip

{\begin{table}[h]
\small
\begin{tabular}{|c|c|c|c|c|c|c|c|c|c|c|c|c|c|c|c|c|c|c|c|}\hline
p &  2 &  3 &  5 &  7 & 11 & 13 &  17 & 19 & 23 & 29 & 31 & 37 &  41 &  43 &  47 & 53 &...& 97 \\
\hline
f(100)&  97 & 48 & 24 & 16 & 9 &  7 &  5 &  5 &  4 &  3 &  3 &  2 &  2 &  2 &  2 &  1 & ... & 1 \\
\hline
f(1000)&  98 & 50 & 25 & 16 & 10 &  6 &  6 &  5 
&  2 &  3 &  4 &  3 &  3 &  3 &  2 &  2 & ... & 1\\ 
\hline
\end{tabular}
\caption{\label{Table1} Prime factor frequencies in the intervals 1 - 100 and 900-1000}
\end{table}}
\medskip

\begin{figure}[h]
\centerline{\epsfig{file=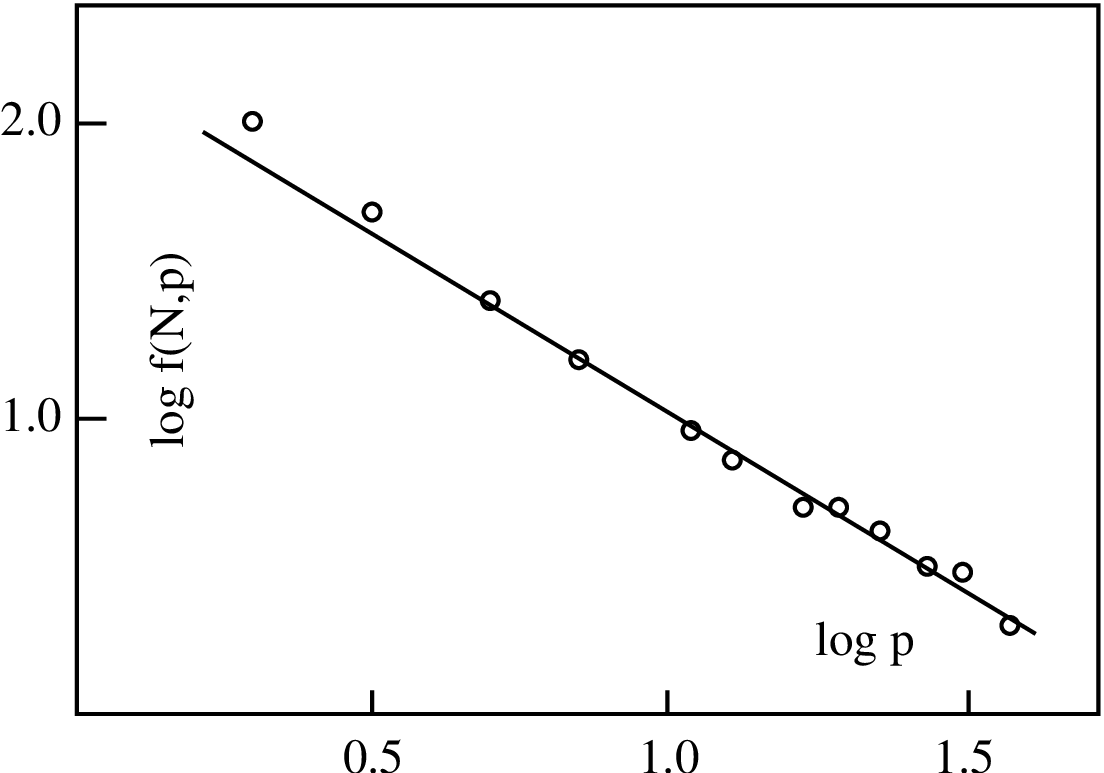,width=8cm}}
\medskip
\centerline{Fig. 5: Prime number frequencies for integers 1-100, compared to
 Zipf's law}
\end{figure}

\medskip

To see if this pattern also continues in another ``text'', we compare the
results for integers between 1 and 100 (text A) with those between 900 and 
1000 (text B); the factorization tables for both texts are given in the 
Appendix. 
In the first case, we have one and two digit integers, in the second three 
digits. 
In the first case, we have 25 direct prime numbers, in the second only 14. 
In the first case, there are four direct prime numbers in the first decade, in
the second there is only one. Thus the two texts indeed are quite disctinct, a
numerical analogue of ``Romeo and Juliet'' vs. ``Moby Dick''. Nevertheless, as 
shown in Table 1, the frequencies of the prime number factors in these two 
texts are indeed essentially the same.

\medskip

So far, we have measured the occurrence frequencies of prime numbers 
empirically, counting occurrences in the specific ``texts'' of integers given 
in the appendix. However, 
the general ocurrence frequency for a prime number $p$ in a large set 
of $N$ integers can in fact be derived analytically from the relation
\be
f(N,p) = N(1/p + 1/p^2 + 1/p^3 +...) \simeq {N\over(p-1)},
\ee 
converging for large $N$. With
\be
\log~f(N,p) \simeq \log N - \log (p-1)
\ee
we obtain a general form of Zipf's law. It is here not found in an analysis
 of empirical measurements, but it is derived as the result of the fixed
distribution
of prime numbers. The replacement of $\log p$ by $\log (p-1)$, as given by the
asymptotic form of eq. (4), makes the fit of
the results essentially perfect,
even for the finite interval 1-100. 
In Fig. 6 we show the logarithmic relation, $\log
f(N,p)$ vs. $\log (p-1)$, and we find indeed the linear Zipf-form.  

\begin{figure}[h]
\centerline{\epsfig{file=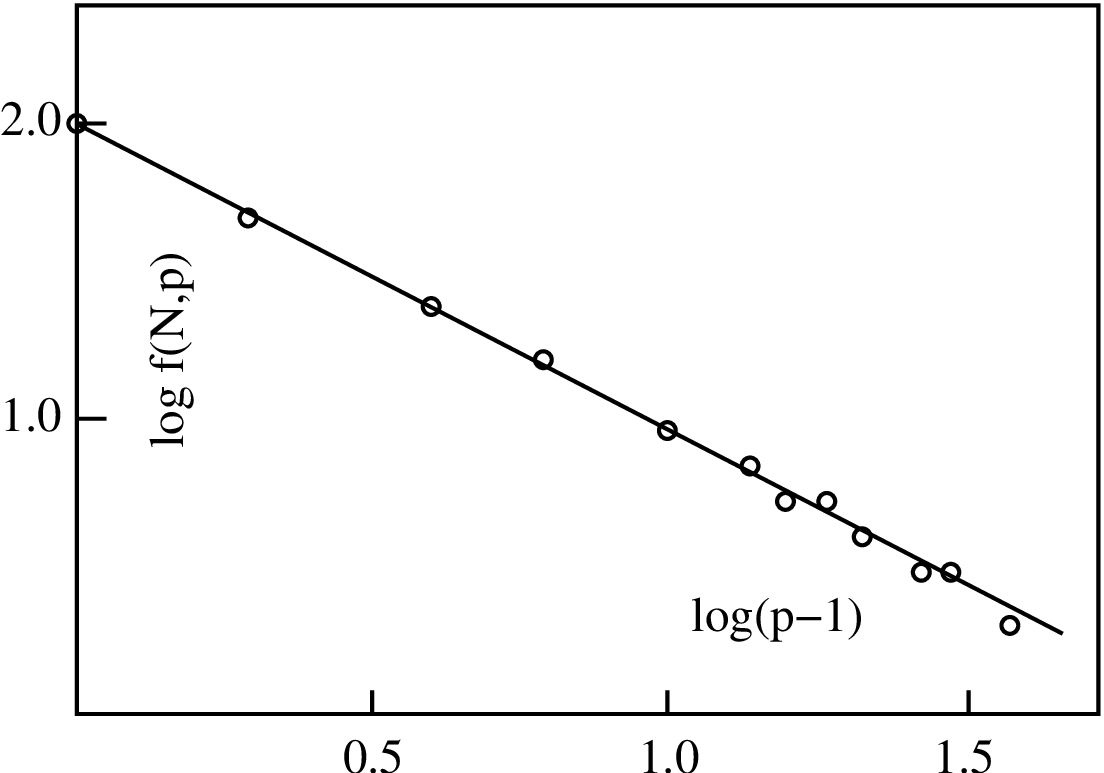,width=8cm}}
\medskip
\centerline{Fig. 6: Prime number frequencies for integers 1-100, compared t0 a
generalized Zipf law}
\end{figure}

Mathematically, the Zipf pattern is a power law for the frequency
in terms of the individual components or events. Let us recall how such
behavior arises in statistical mechanics and consider the perhaps simplest
case there, the two-dimensional Ising model. We have $N$ spins $s_i=\pm1,
i=1,2,...N$  on a two-dimensional lattice. The correlation function 
$F(r_{ij},T)$ at temperature $T$, with $r_{ij}=s_i-s_j$ for the separation 
distance between spins $i$ and $j$, is given by
\be
F(r_{ij},T) \sim {\exp{-[r_{ij}/\lambda(T)]} \over r_{ij}},
\ee
where $\lambda(T)$ denotes the correlation length at temperature $T$, as 
determined by the dynamics
of the system. Thus the correlation depends in general on two parameters: the
separation distance $r_{ij}$ and through the correlation length on the
temperature. The dependence is dominantly exponential, and hence the
correlation function is not scale-invariant:  $x F(xr,T)\not=F(r,T)$. The
situation changes only at the critical point, $T=T_c$, where the correlation
length diverges: $\lambda(T_c)=\infty$. The correlation function now becomes
power-like,
\be
 F(r_{ij},T_c) \sim {1 \over r_{ij}},
\ee
and hence scale-invariant,
\be
x F(x r_{ij}, T_c) = F(r_{ij},T_c).
\ee
We thus find that in such systems power-law behavior is a sign of
criticality. In the non-critical regime, the correlation function exhibits 
an exponential fall-off for large distances $r$, determined by $\lambda$, 
while at the critical temperature, it becomes a power-law in $r$.
 
\vskip1cm

\centerline{\bf \large Acknowledgement:}

\bigskip

It is a pleasure to thank Andrew Granville (Montreal) for helpful comments.

\newpage

~\vskip2cm

\begin{figure}[h]
\centerline{\epsfig{file=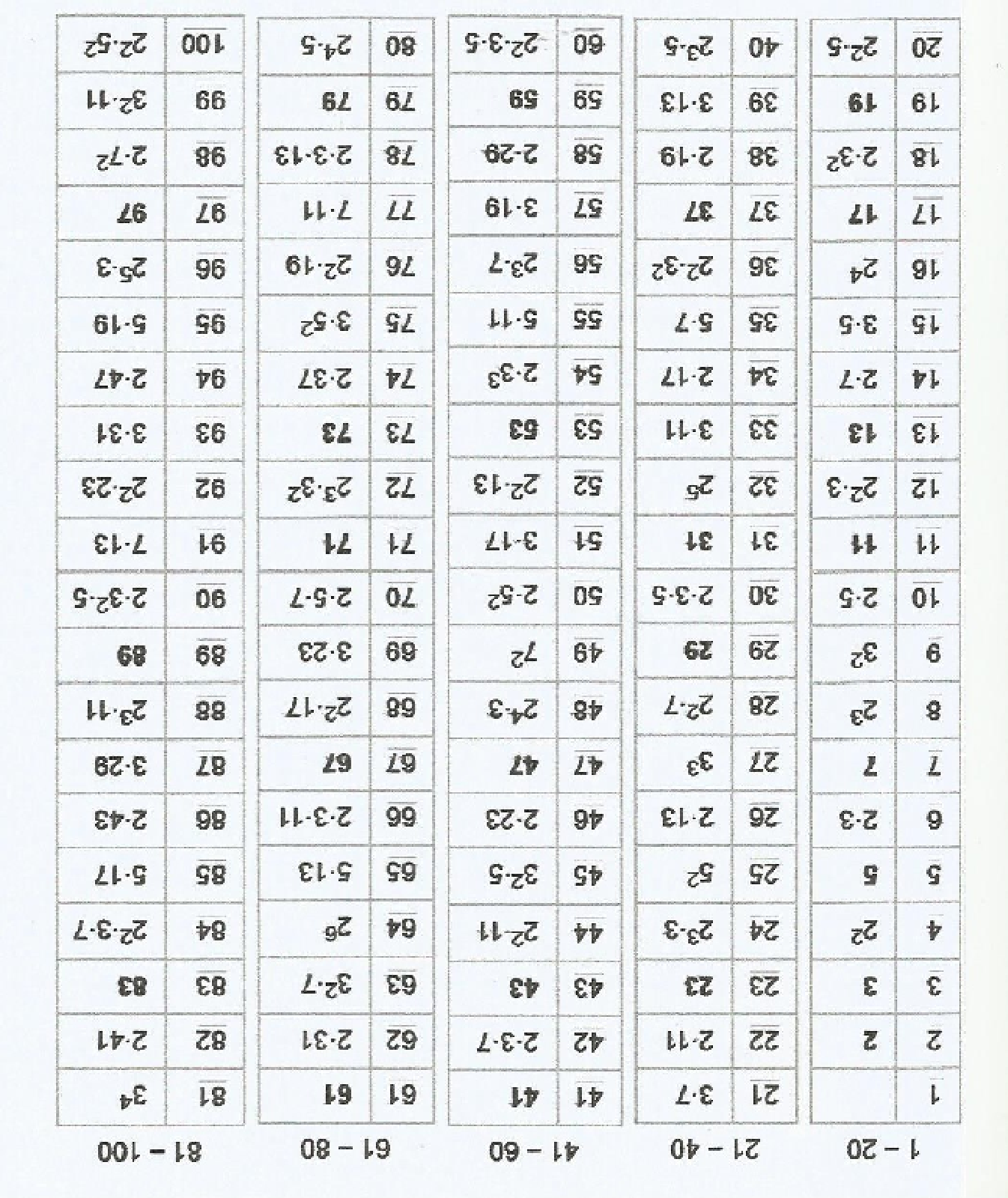,width=14cm,angle=-180}}
\bigskip
\centerline{Appendix A: Prime factor decomposition of integers 1 - 100}
\end{figure}

\bigskip

\centerline{\sl From Wikipedia: Table of prime factors}

\newpage

~\vskip2cm
\begin{figure}[h]
\centerline{\epsfig{file=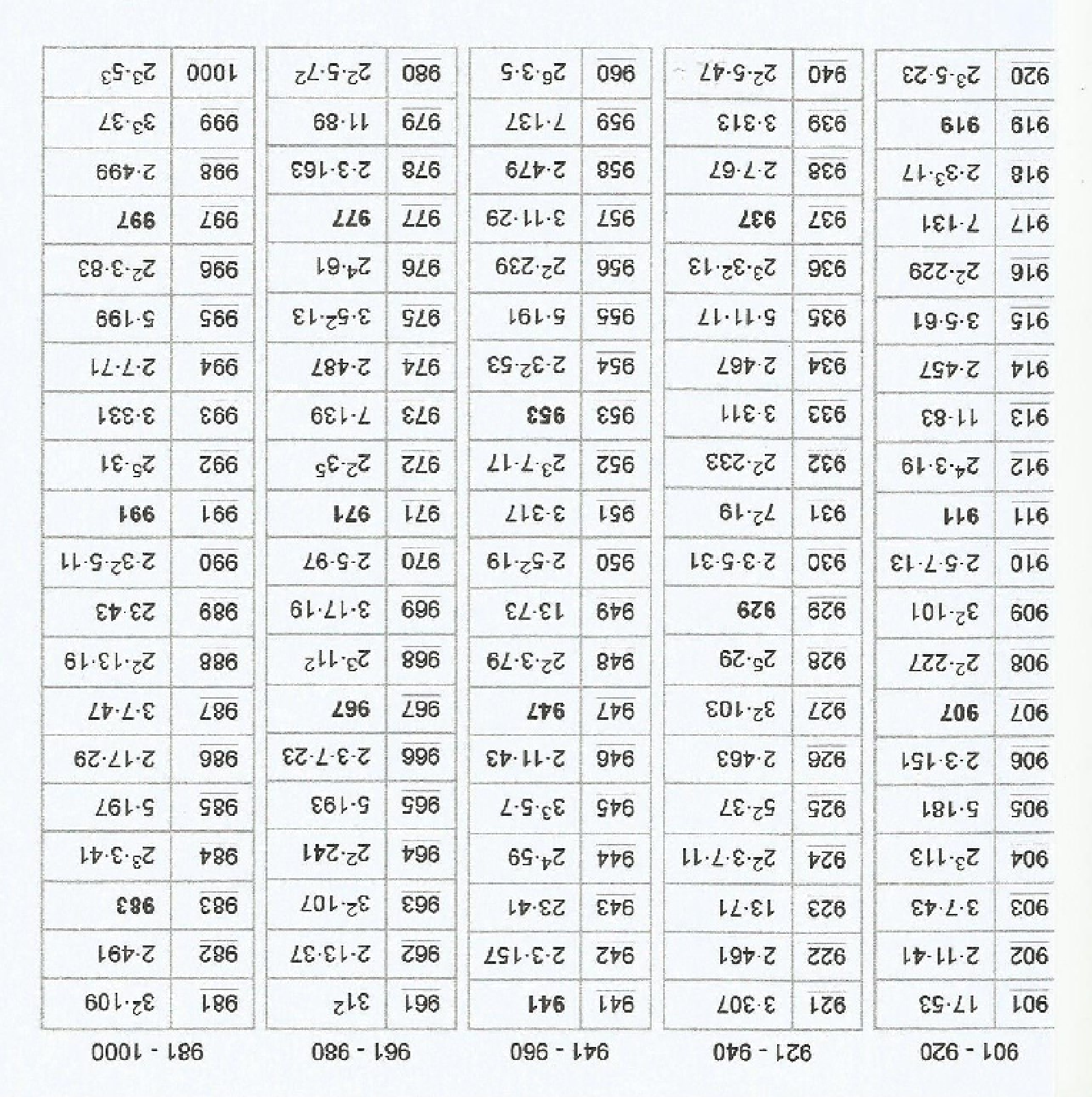,width=16cm,angle=-180}}
\bigskip
\centerline{Appendix B: Prime factor decomposition of integers 900 - 1000}
\end{figure}

\bigskip

\centerline{\sl From Wikipedia: Table of prime factors}

\end{document}